# Female Student Population at Kabul University Before the 2021 Ban: Trends, Gender Parity, and Faculty-Level Dynamics

Dr. Jawid Ahmad Baktash

Technical University of Munich

Dr. Mursal Dawodi

Technical University of Munich

**Abstract**

For nearly two decades after 2001, Afghanistan's higher education sector expanded rapidly, with Kabul University serving as a central site of women's academic participation. Drawing on administrative records of student populations from 2016-2019 (Islamic calendar 1395-1398), this study examines gender distributions across shifts, faculties, and departments, with particular attention to STEM versus non-STEM fields. At Kabul University, the morning shift refers to the main daytime cohort (including some midday classes), while the evening shift is a separate program with its own classes and students; the two cohorts are administratively and academically distinct. Results show steady growth in the overall female student population, but with marked disparities between morning and evening shifts. Women were concentrated in non-STEM and "socially acceptable" disciplines such as literature, law, and psychology, while within STEM they were relatively well represented in the life sciences but remained significantly underrepresented in technical fields such as engineering, computer science, and physics. Gender parity improved modestly across most faculties, yet the Gender Parity Index (GPI) rarely approached 1.0, and the STEM GPI consistently remained below 0.5. These findings highlight both progress and persistent structural inequalities, documenting a critical historical benchmark before the 2021 ban on women's university education.

Keywords: Afghanistan, higher education, female students, gender equity, Kabul University, STEM, Gender Parity Index, Taliban ban

## 1. INTRODUCTION

Following the fall of the Taliban in 2001, Afghanistan entered a period of rapid reconstruction of its higher education system. With international donor support and government reforms, access expanded rapidly, and women became increasingly visible on university campuses. Kabul University, founded in 1931, is Afghanistan's oldest and most influential institution of higher education (Rubin, 2013). As the country's flagship university, it has shaped intellectual, political, and cultural life for nearly a century. Women have long played a role in this history, but their access to higher education has fluctuated dramatically depending on political regimes and social conditions. At Kabul University, female students gained strong representation in pharmacy, law, and the social sciences (Rahimi, 2019; World Bank, 2020). Yet progress remained uneven. Structural barriers, such as shortages of dormitories and female faculty, financial constraints, and cultural norms, restricted access for many women, particularly in technical fields such as engineering and information technology (Esmaeily et al., 2010; Ahmed-Ghosh, 2003; Bukhari, 2023).

These fragile gains were reversed after the Taliban's return to power in August 2021. By December 2022, a nationwide ban on female university attendance excluded more than 110,000 women from higher education (UNESCO GEM, 2023), erasing two decades of advancement. Against this backdrop, the years 2016 (1395) to 2019 (1398) hold particular significance as the last verifiable record of women's participation in Afghan higher education before their exclusion.

This paper analyzes official Kabul University student population data from 2016–2019 to examine how gender distributions shifted across faculties, departments, and study shifts[1]. Specifically, it asks how the overall gender composition changed in these years, which faculties achieved more balanced representation, and how women's participation in life sciences compared to their near absence in technical STEM fields. By addressing these questions, the study not only documents a critical moment in Afghan higher education but also contributes to broader debates on gender, education, and fragility in conflict-affected contexts.

2. **Literature Review**

Research on women's higher education in Afghanistan highlights both the remarkable expansion that occurred after 2001 and the devastating reversal that followed the Taliban's return in 2021. In the early 2000s, fewer than 6,000 students were enrolled nationwide. By 2020, that number had grown to nearly 400,000, including more than 110,000 women across 39 public universities and 129 private institutions (Rahimi, 2019; World Bank, 2020). By 2018, women constituted almost one-third of all university students, a remarkable shift compared to previous decades.

Despite this progress, women's participation remained highly uneven. Female students were disproportionately concentrated in medicine, education, and the social sciences, while their representation in engineering, ICT, and other technical STEM fields remained minimal (Esmaeily et al., 2010). Barriers to entry included inadequate infrastructure—particularly the shortage of women's dormitories that limited access for students from rural areas—along with cultural norms that discouraged women from pursuing technical disciplines and security concerns that disproportionately affected them (Ahmed-Ghosh, 2003; Bukhari, 2023). Structural inequalities in admissions and scholarships also reinforced gender gaps, though affirmative action policies such as gender quotas in the 2010s modestly increased women's participation in some faculties (Rahimi, 2019).

At the global level, research consistently emphasizes the fragility of women's educational gains in conflict-affected settings. Studies show that women's education is often among the first casualties of political instability, yet it remains one of the most critical foundations for reconstruction and peacebuilding (Stromquist, 2015; Kirk, 2008; Dryden-Peterson, 2020). Comparative regional evidence underscores Afghanistan's lag. By 2020, women's tertiary enrollment worldwide had surpassed men's—43% versus 37%—with South Asia also showing steady increases (UNESCO, 2022; IFC, 2025). However, UNESCO's 2023 Gender Report cautions that these global gains are uneven: women remain significantly underrepresented in STEM and leadership positions, especially in low-income and conflict-affected contexts (UNESCO GEM, 2023).

A further nuance, often overlooked in the Afghan literature, is the internal division within STEM. While women are generally underrepresented in STEM fields, international evidence shows that life sciences (biology, pharmacy, chemistry) often approach parity or even female majority, whereas technical STEM (engineering, ICT, mathematics, physics) remain persistently male-dominated (Charlesworth & Banaji, 2019). Scholarship focusing on Afghanistan has rarely examined this distinction. Existing studies treat STEM as a single block, thereby obscuring the areas where women have been successful.

The theoretical framing offered by Stromquist (2015) and Unterhalter (2017) is instructive: gender equity in higher education should not be measured by numbers alone, but by the sustainability and resilience of gains under fragile political conditions. Afghanistan represents a stark case. Between 2001 and 2021, donor support, quotas, and advocacy contributed to visible progress, yet these achievements rested on precarious foundations. The Taliban's December 2022 decree banning women from universities erased two decades of advancement and excluded more than 110,000 women from higher education.

---

[1] At Kabul University, the student body is organized into two distinct cohorts: the morning shift (also referred to as the day shift) and the evening shift. Morning-shift students constitute the main cohort and attend standard daytime classes, which may occasionally extend into midday hours in some faculties, though they remain part of the same cohort. Evening-shift students, by contrast, are enrolled in a separate program with their own classes, instructors, and administrative structure. Students enrolled in one shift cannot attend courses or examinations from the other shift. These cohorts are therefore academically and administratively distinct, and the analysis in this study treats them as separate populations.

In this context, the years 2016–2019 represent not only a period of growth but also the last verifiable record of women's active participation in Afghan higher education. Kabul University, as the country's flagship institution, offers a critical case study. By disaggregating student population data across faculties, departments, and shifts, this study contributes to the literature by documenting both progress and persistent inequalities, while addressing the overlooked distinction between life sciences and technical STEM.

3. **Methodology**

This study adopts a mixed-methods design on quantitative analysis of Kabul University's student population. The focus is on the years 1395–1398 (2016–2019), representing the last verifiable dataset before the Taliban's December 2022 ban on women's higher education.

The dataset was drawn from official administrative records, which reported the total number of registered students disaggregated by gender, year, semester, faculty, department, and study shift (morning or evening). Data availability varied across years: in 1395 (2016), only aggregated morning/evening totals were available; in 1396 (2017), fall semester data were available by shift; and in 1397–1398 (2018–2019), records included both spring and fall semesters. To maintain comparability and avoid double-counting, only fall semester data were used in the analysis.

The distinction between morning and evening shifts at Kabul University is not merely administrative but also socially significant. Morning shifts are merit-based and tuition-free, enrolling recent high school graduates from diverse socio-economic backgrounds. Evening shifts are fee-paying programs, often attended by older or working students. Because evening classes occur later in the day, they are less culturally acceptable for women and are perceived as less safe, leading to markedly lower female participation. Analyzing both shifts separately is therefore essential to understanding gendered access.

The analysis was guided by three objectives: (i) to measure overall changes in gender composition at the university between 2016 and 2019, (ii) to compare gender distributions across shifts, faculties, and departments, and (iii) to examine disparities between STEM and non-STEM fields, with particular attention to the divide between life sciences and technical disciplines.

**4. Data Analysis**

Data processing involved several steps. Missing values were corrected, and categorical variables (faculty, department, and shift) were standardized. Morning and evening sessions were analyzed separately, since evening programs were not uniformly available across faculties and typically had lower female participation. Faculties were classified into broad academic domains: non-STEM (e.g., social sciences, law, sharia, literature), STEM-life sciences (biology, chemistry, pharmacy), and STEM-technical (engineering, ICT, mathematics, physics). This distinction allowed for a more nuanced analysis of women's representation in science-related disciplines, reflecting global literature on gender disparities within STEM.

Several key variables were constructed. Female share (%) was defined as:

$$FS(\%) = \frac{F}{F+M} \times 100 \qquad (1)$$

Where FS = female share, F= number of female students and M= number of male students

The Gender Parity Index (GPI) was calculated as:

$$\text{GPI} = \frac{F}{M} \qquad (2)$$

At the university-wide level, GPI was computed using totals across all faculties, rather than averaging faculty-level ratios, to avoid distortions caused by small-group variation.

To test whether gender distributions were independent of institutional structures, chi-square tests of independence were calculated, alongside Cramér's V as an effect size measure:

$$V = \sqrt{\frac{\chi^2}{n \cdot \min(r-1, c-1)}} \qquad (3)$$

where $x^2$= chi-square statistic, n is the total number of students in the table, r the number of rows, and c the number of columns. Cramér's V values of 0.1, 0.3, and 0.5 are conventionally interpreted as small, moderate, and large effects, respectively.

The analysis proceeded in three complementary stages. First, descriptive statistics were calculated to capture totals and gender shares at different levels of aggregation (university-wide, shifts, faculties, departments). Second, a range of visualizations was used, line plots for overall trends, stacked bar charts for shifts and faculties, treemaps for the distribution of female students, bubble plots linking female share to faculty size, and heatmaps for GPI by faculty. Third, robustness checks were applied to confirm validity. Excluding 1395/2016 data produced no changes in trends, confirming that treating it as fall-equivalent did not bias results. Different GPI calculation methods yielded consistent outcomes, and restricting analysis to faculties offering both morning and evening shifts still showed disproportionate female presence in morning programs.

All analyses were performed in Python using pandas for data manipulation, matplotlib for visualization, and SciPy for statistical testing.

## 5. Results

The analysis of Kabul University's student population between 2016 and 2019 (1395–1398) reveals clear gains in women's participation alongside persistent structural inequalities. At the university level, the female share rose from about 41% in 2016 to nearly 47% in 2019, with absolute numbers increasing from roughly 5,700 to 8,700 while male enrollment fluctuated slightly. This narrowed the gender gap, and the Gender Parity Index (GPI) improved from 0.70 to 0.89 (Figure 1).

Participation patterns differed markedly between study shifts. In morning programs, women comprised 40–50% of students in most faculties, and their numbers increased from about 5,200 to more than 8,200. Evening programs remained strongly male-dominated, enrolling only 500–600 women annually and rarely exceeding one-quarter women in any faculty (Figure 2). Faculty-level distributions reinforce this pattern: morning programs show relatively balanced representation in a number of faculties (Figure 3a), whereas evening programs are overwhelmingly male (Figure 3b). Statistical tests (reported in the methods) confirm that these differences are systematic rather than random.

Faculty-level variation was pronounced. By 2019, several faculties, including Foreign Literature, National Literature, Law and Political Sciences, Social Sciences, Psychology and Educational Sciences, and Pharmacy, recorded female shares above 45–55%, in some instances surpassing parity. In contrast, Engineering, ICT, Computer Science, and Physics consistently reported female shares below 20%, and Sharia remained below 30%. Faculties such as Economics, Chemistry, Geology, and Environment fell into an intermediate band (25–40%), with gradual gains but no approach to parity (Figures 4–5).

The distribution of women was concentrated in a limited set of large faculties. The treemap (Figure 6) shows that Foreign Literature, National Literature, Law and Political Sciences, Social Sciences, Psychology and Educational Sciences, and Pharmacy account for most female students. The bubble plot (Figure 7) adds that large faculties with high female shares absorbed the bulk of women, while comparably large but male-dominated faculties, such as Engineering or other sizable technical fields, remained relatively closed.

Aggregating by domain, the divide between STEM and non-STEM is clear (Figure 8). By 2019, women represented more than half of students in non-STEM fields. Within STEM, life sciences (Biology, Chemistry, Pharmacy) approached parity and sometimes reported female majorities, while technical STEM fields (Engineering, ICT, Mathematics, Physics) remained overwhelmingly male, with women rarely

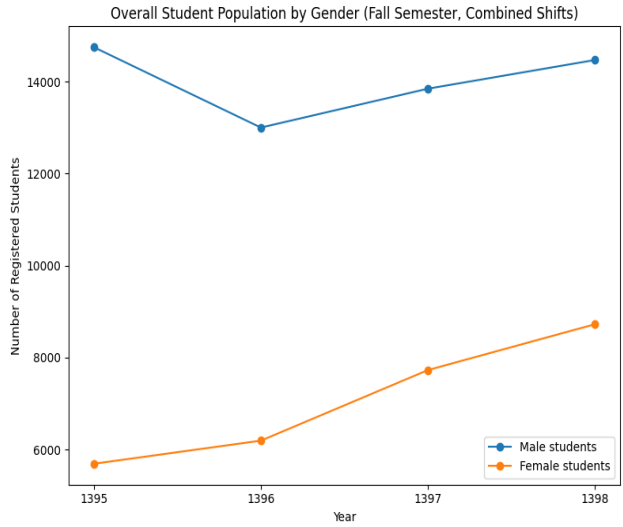

Figure 1. Overall Student Population by Gender (Fall Semester, Combined Shifts), 2016–2019

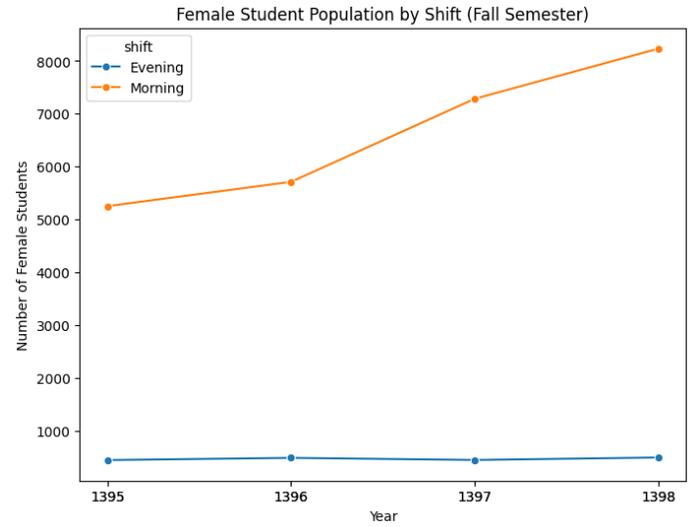

Figure 2. Female Student Population by Shift (Fall Semester), 2016–2019.

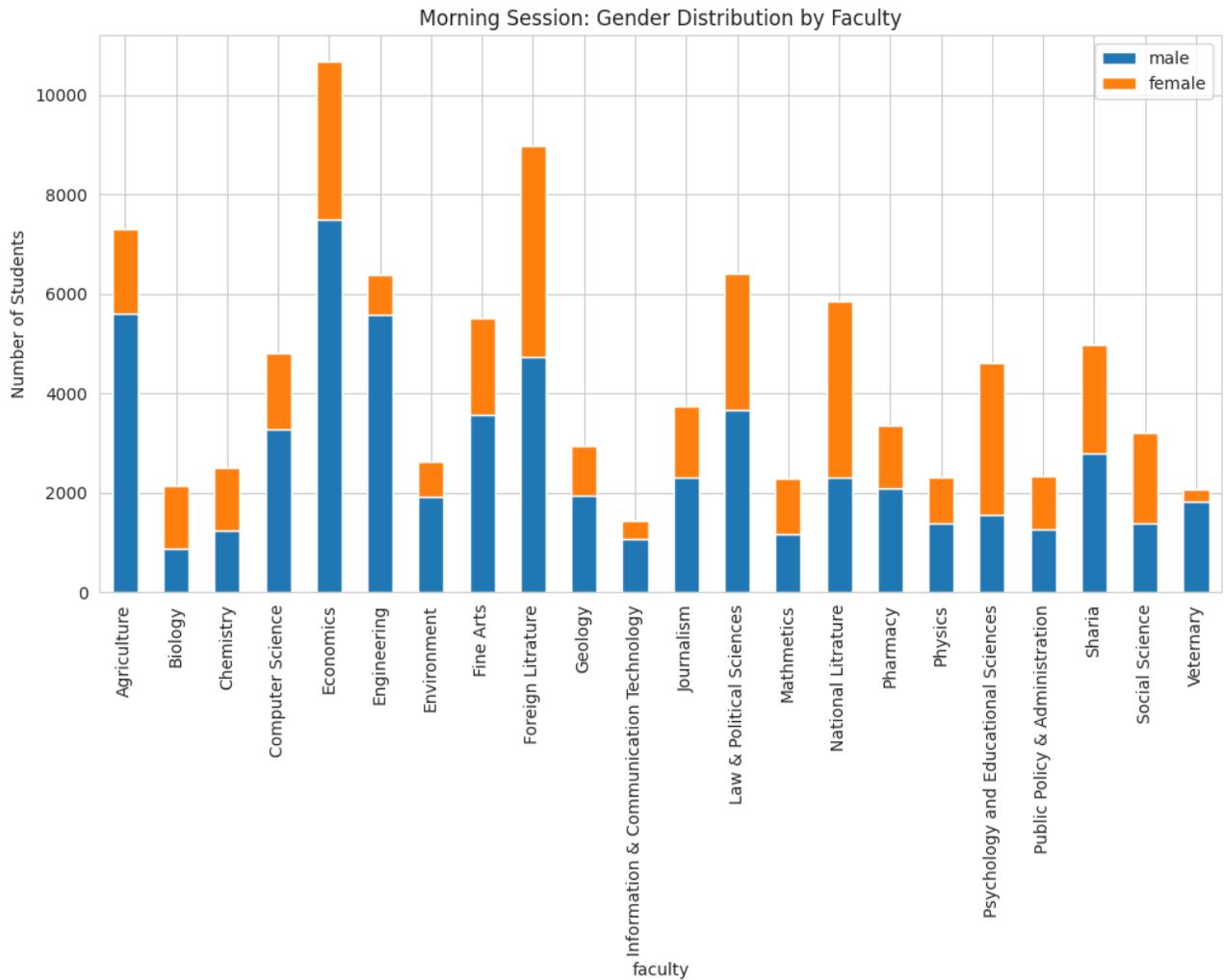

Figure 3a. Gender Distribution by Faculty and Morning Shift

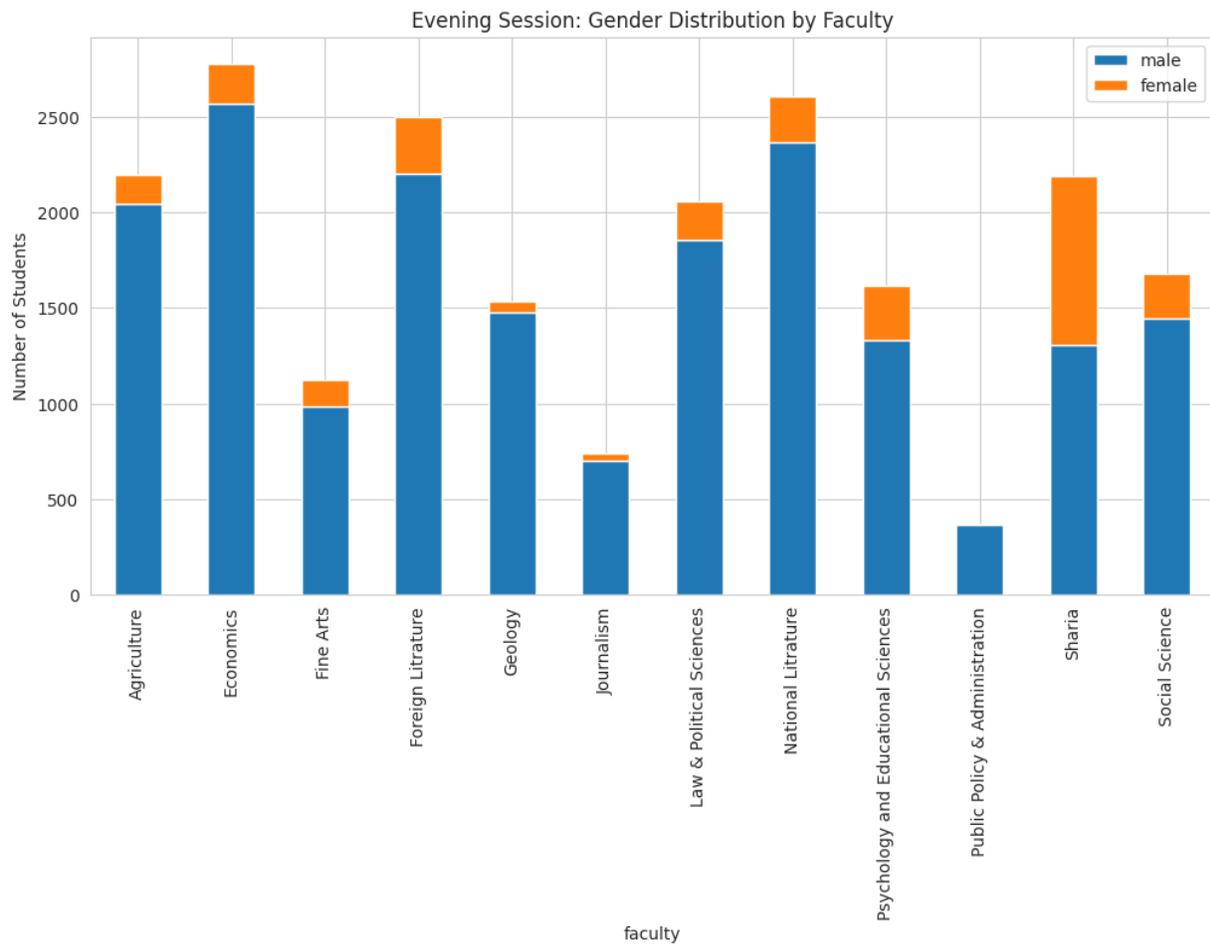

Figure 3b. Gender Distribution by Faculty and Evening Shift

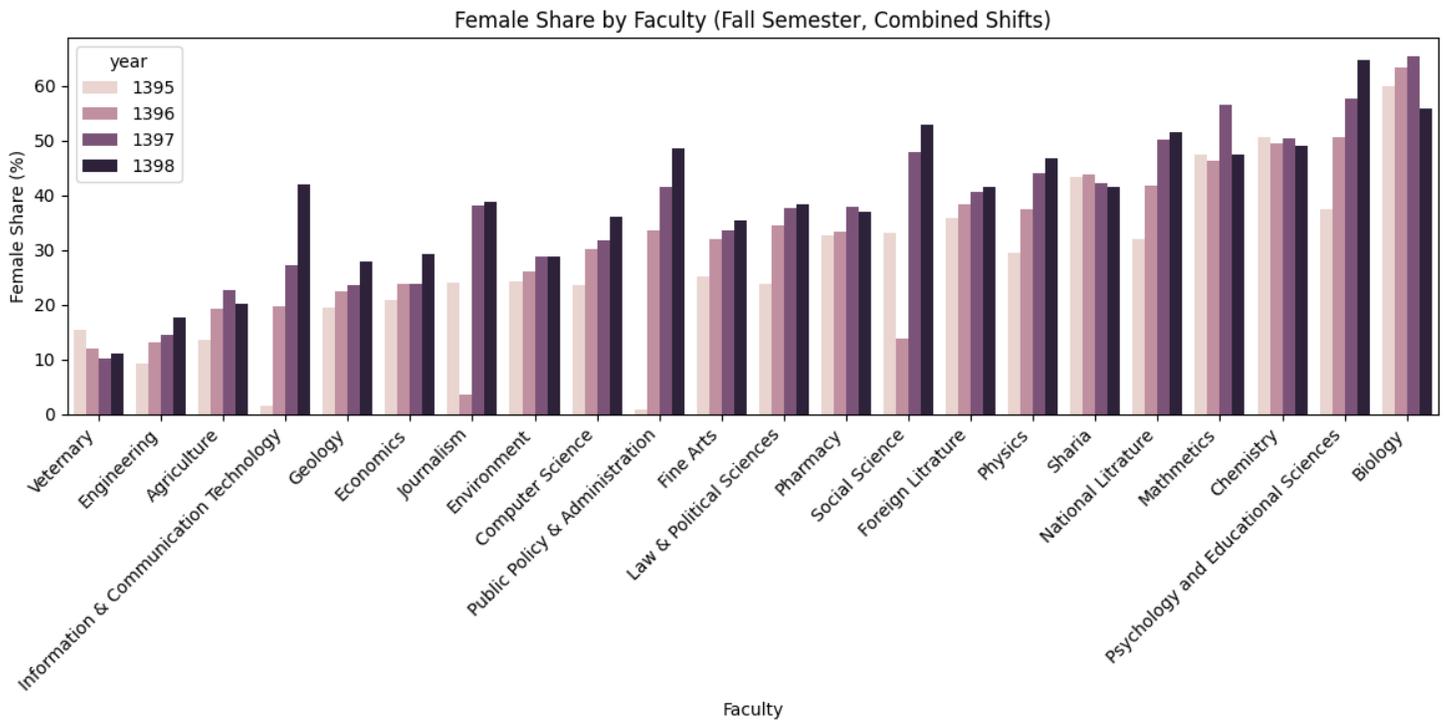

Figure 4. Female Share by Faculty (Fall Semester, Combined Shifts)

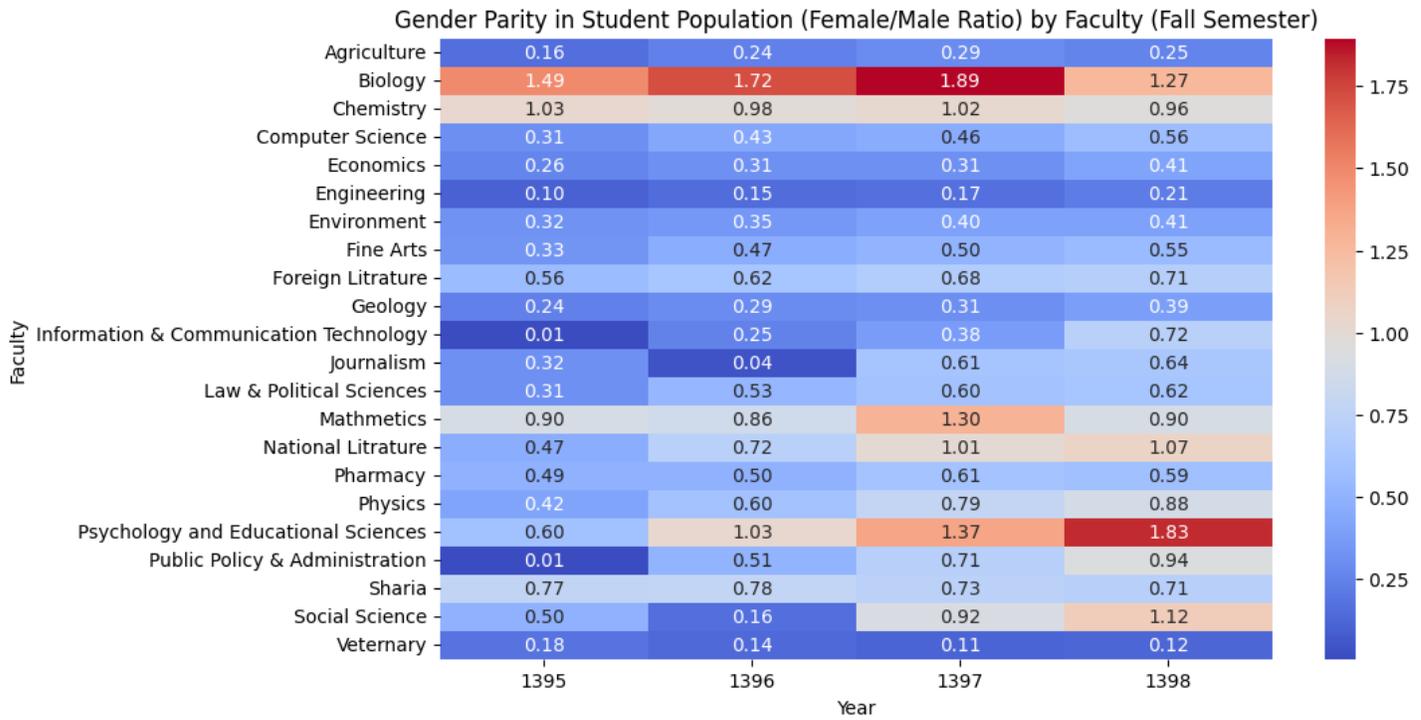

Figure 5. Gender Parity Index (Female-to-Male Ratio) by Faculty (Fall Semester, Combined Shifts)

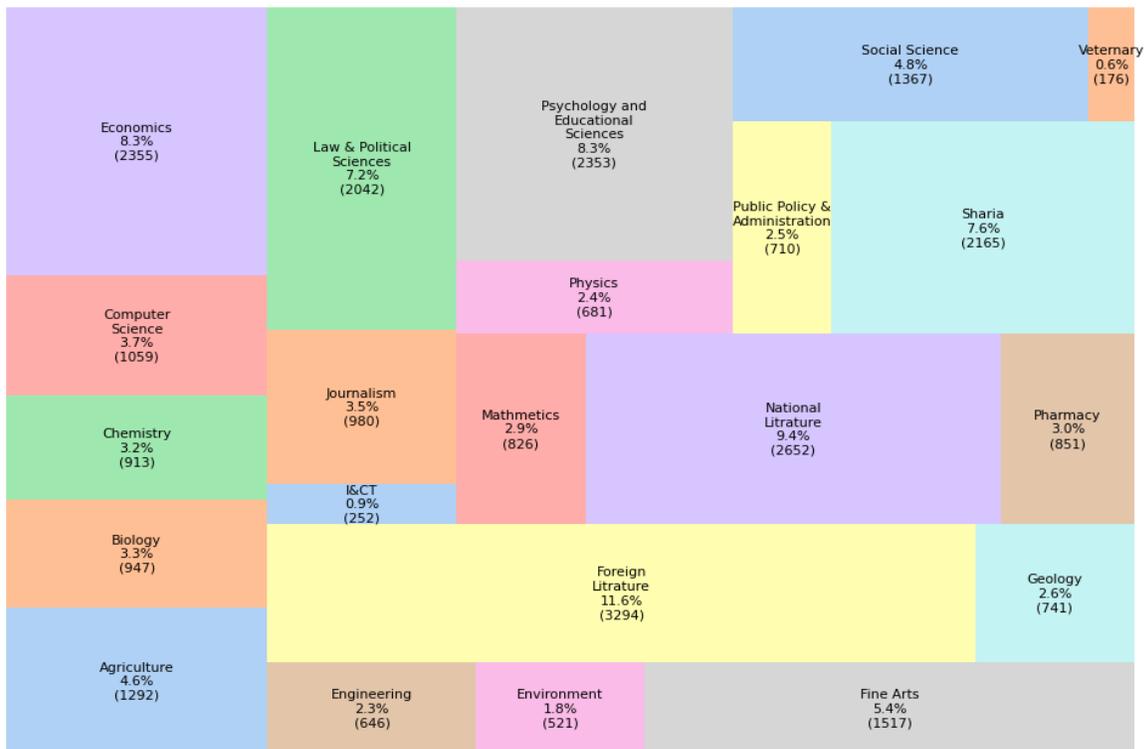

Figure 6. Distribution of Female Students Across Faculties (Fall Semester, Combined Shifts)

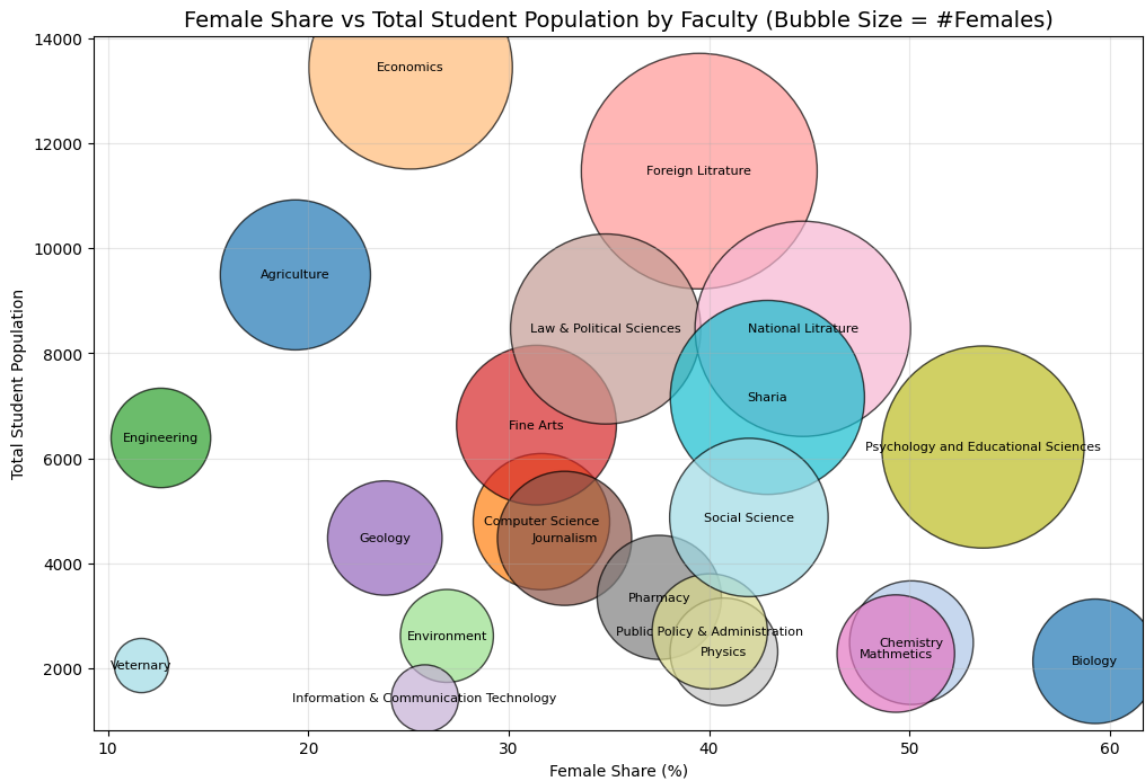

Figure 7. Female Share vs. Total Student Population by Faculty (Fall Semester, Combined Shifts)

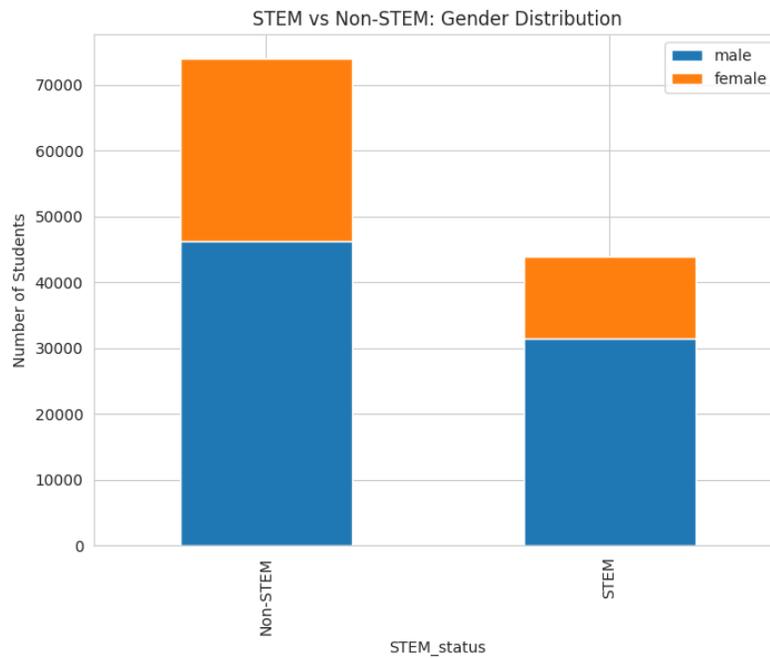

Figure 8. STEM vs. Non-STEM Gender Distribution (Fall Semester, Combined Shifts).

exceeding 10–15%. The GPI trend for the university as a whole rose from 0.70 to 0.89 across the period (Figure 9), but disaggregation shows that gains concentrated in areas such as Foreign Literature, Psychology and Educational Sciences, and Biology, while technical STEM and Sharia remained far below parity. Department-level visualization (Figure 10) confirms that humanities and life-science departments often approached balance, whereas technical departments, including Civil and Electrical Engineering, Computer Science, and Physics, remained heavily male.

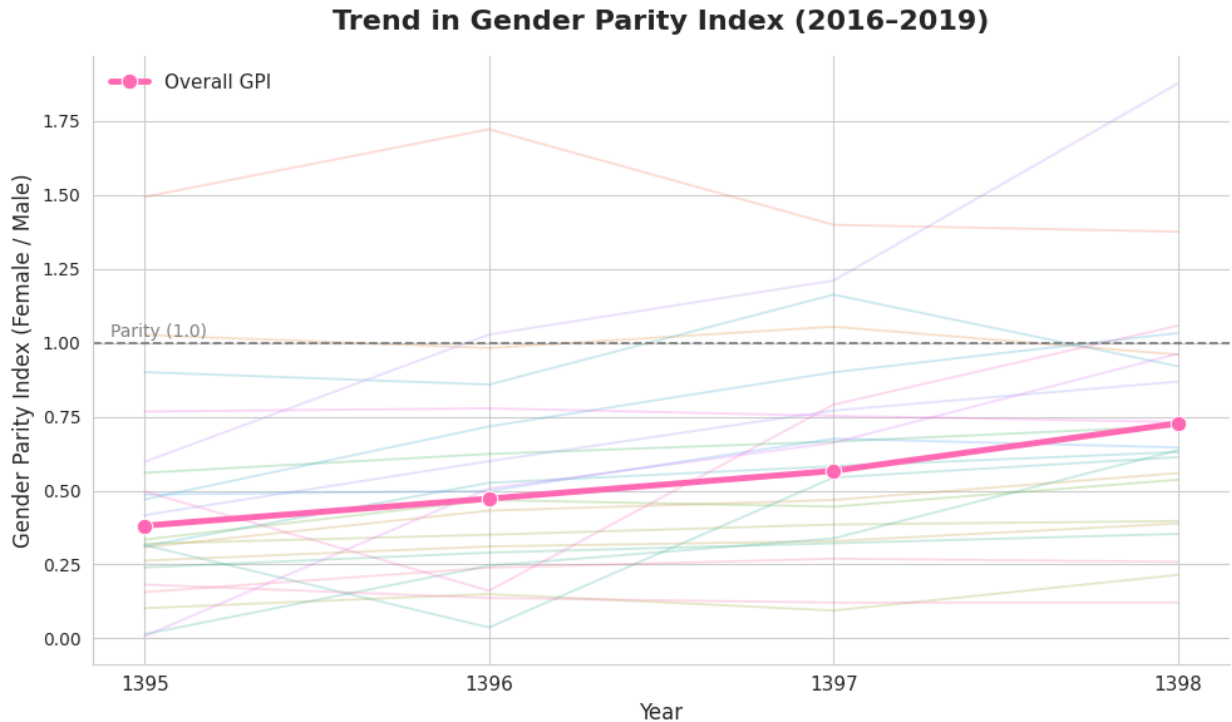

Figure 9. Gender Parity Index Trends (Fall Semester, Combined Shifts), 2016–2019

Taken together, these results indicate that Kabul University moved closer to overall gender balance by 2019, but progress was unevenly distributed. Gains accrued primarily in morning programs, non-STEM disciplines, and life sciences, while evening shifts, technical STEM faculties, and religious faculties remained disproportionately male.

## 6. Discussion

The findings from Kabul University between 2016 and 2019 reveal that women's rising participation in higher education was shaped less by a uniform expansion of opportunity than by a series of layered and intersecting inequalities. The overall increase in women's share of enrollment suggests progress, yet the structures of access determined where and how that progress occurred, creating distinct patterns of inclusion and exclusion.

One clear layer of inequality emerges in the contrast between morning and evening programs. Morning shifts, as tuition-free and merit-based programs, were the most accessible route into higher education for women and therefore absorbed nearly all of the female enrollment growth. By contrast, evening programs, which required fees and operated at less socially acceptable hours, remained overwhelmingly male. This structural divide meant that while women were gaining visibility on campus, their opportunities were circumscribed by program type: economic barriers and cultural restrictions effectively excluded them from one entire track of higher education.

These institutional divisions reinforced disciplinary segregation. Because the evening programs were heavily concentrated in faculties such as Economics, Law, and Engineering, women's limited access to them deepened their concentration in morning programs where "socially acceptable" disciplines — literature, psychology, pharmacy — were more prominent. In this way, the exclusion from evening shifts was not just about time of day; it was a mechanism that indirectly steered women into certain fields while shutting them out of others.

Within STEM, the internal divide between life sciences and technical fields further magnified the unevenness of progress. The relative success of women in biology, chemistry, and pharmacy shows that female participation in science was possible where institutional support and cultural legitimacy aligned. But their near-absence from engineering, computer science, and physics highlights the persistence of gendered hierarchies that equated technical expertise with masculinity. Together, these patterns meant that even as women approached parity overall, the symbolic and material rewards of higher education — such as entry into technical professions, leadership roles, and high-status employment — remained disproportionately male.

Seen in combination, these findings suggest that Afghanistan's post-2001 expansion of higher education produced numerical gains without dismantling the structural foundations of inequality. Women could enter universities in growing numbers, but the pathways open to them were narrow, shaped by a convergence of program design (morning vs. evening), disciplinary hierarchies (life sciences vs. technical STEM), and social norms (acceptable vs. unacceptable fields). When these fragile gains were reversed after 2021, it underscored that access alone is insufficient without institutional and cultural reforms that embed equity more deeply into the fabric of higher education.

From this perspective, the data do more than document the past. They show how inequality operates through multiple layers — program structure, disciplinary boundaries, and cultural legitimacy — and how reforms must address all three if future gains are to be both inclusive and sustainable.

6.1 Limitations and Future Research

This study has several limitations that should be considered when interpreting the findings. First, the analysis draws on administrative records from a single institution, Kabul University. As the country's flagship university, it provides a strong case study, but the results may not fully capture patterns in smaller public institutions or the rapidly growing private sector. Second, the dataset records aggregate enrollments rather than unique individuals, meaning that continuing students appear in multiple semesters and progression cannot be measured directly. Third, the absence of socio-economic, geographic, and ethnic indicators limits our ability to understand why some women were able to participate while others were excluded. Finally, although qualitative insights from former Ministry of Higher Education officials add policy context, the absence of direct student perspectives leaves important lived experiences unexplored.

Future research should expand the scope of analysis to multiple universities across Afghanistan, including private institutions, in order to assess whether Kabul University was representative or exceptional. Incorporating socio-economic and geographic data would make it possible to identify which groups of women gained or were denied access. Longitudinal data that follow individual students would allow researchers to measure persistence, graduation, and career outcomes, while alumni studies could reveal the broader social and economic effects of higher education. If women are eventually readmitted to Afghan universities, systematic monitoring will be essential to determine whether the structural patterns identified here, such as concentration in non-STEM and exclusion from evening programs, persist or change under new conditions.

**7. Conclusion**

This study has examined gendered enrollment patterns at Kabul University during 2016–2019, offering the most complete record of women's participation before their exclusion from higher education. The findings show that women's share of the student body approached parity by 2019, with several faculties achieving balance or even female majorities. At the same time, participation remained uneven, as technical STEM fields and evening programs largely excluded women, reflecting structural, cultural, and institutional barriers.

The significance of this period lies not only in documenting progress but also in exposing the fragility of those gains. The sharp reversal in 2021 demonstrated how quickly opportunities can collapse when they are not supported by durable systems of equity and governance.

Looking ahead, restoring women's access will be only the first step. Sustainable reform will require embedding gender equity into higher-education governance so that progress is resilient to political shocks. Priorities include investing in women's dormitories and safe housing, recruiting and retaining female faculty, creating targeted scholarships and mentorship pathways in technical STEM, and broadening the cultural legitimacy of women's participation across all fields of study.

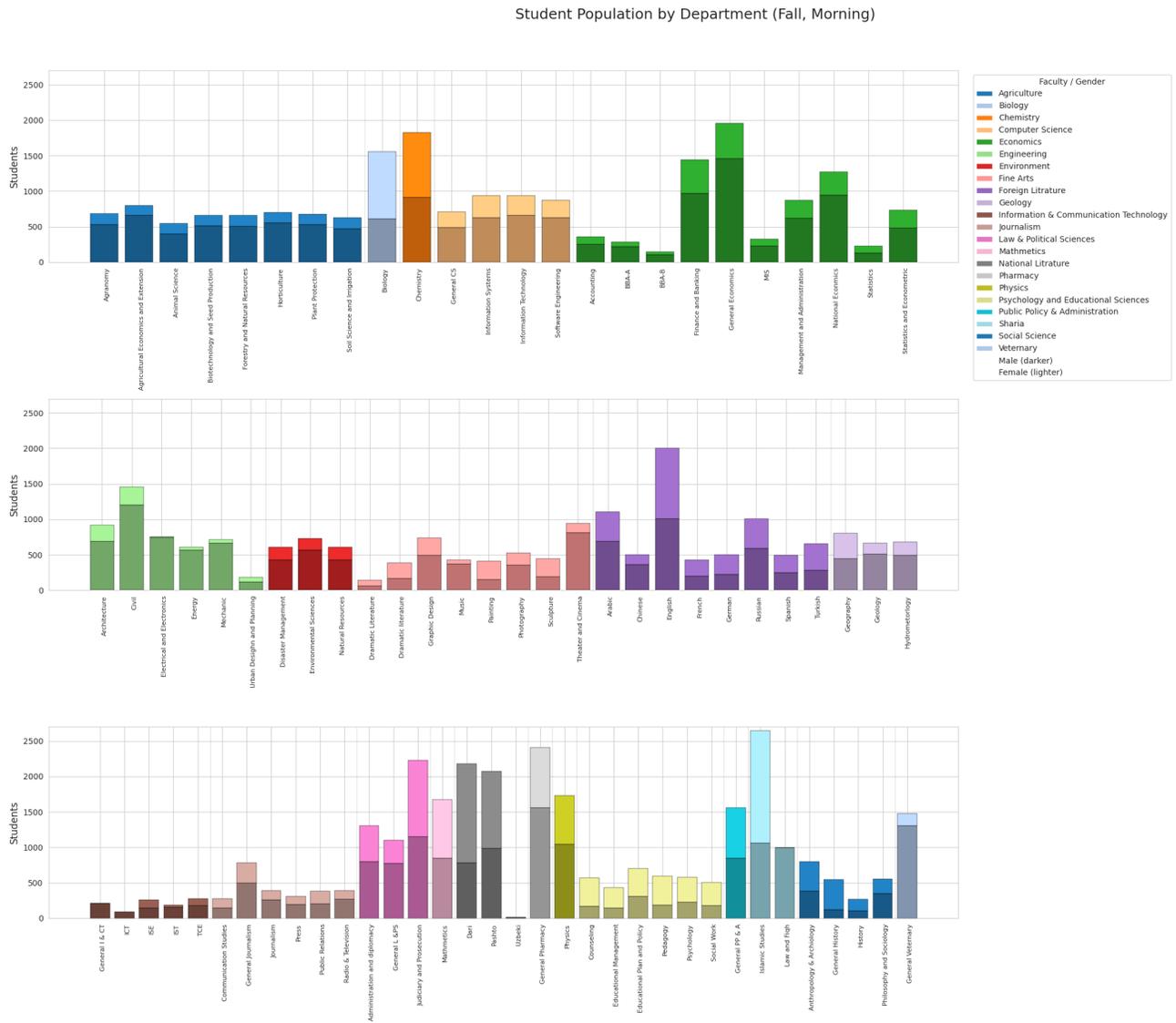

Figure 10. Gender Distribution by Department

By situating Kabul University's experience within both national and global debates, this study provides a historical benchmark and a forward-looking framework. The lessons drawn here can help guide the design of more inclusive, equitable, and resilient institutions when Afghan women are once again able to pursue higher education.